\documentclass[aps,pra,twocolumn,superscriptaddress]{revtex4-2}
\usepackage[T1]{fontenc}
\usepackage[utf8]{inputenc}
\usepackage{amsfonts}
\usepackage{amsmath}
\usepackage{amssymb}
\usepackage{amsthm}
\usepackage{esint}
\usepackage{graphicx}
\usepackage{bm}
\usepackage{color}
\usepackage{xcolor}
\usepackage{mathrsfs}
\usepackage[colorlinks,bookmarks=true,citecolor=blue,linkcolor=red,urlcolor=blue]{hyperref}
\usepackage{appendix}
\usepackage{float}
\usepackage[export]{adjustbox}
\usepackage{ulem}
\usepackage{babel}
\newcommand{\RNum}[1]{\uppercase\expandafter{\romannumeral #1\relax}}

\newcommand{\beq}{\begin{eqnarray} }
\newcommand{\eeq}{\end{eqnarray} }
\newcommand{\Beq}{\begin{eqnarray*} }
\newcommand{\Eeq}{\end{eqnarray*} }

\newcommand{\re}{\mathrm{Re}}

\newcommand{\luvl}{\mathcal{L}}
\newcommand{\luvlD}{\mathcal{L}^\dagger}

\newcommand{\SecIIID}{\ref{SecIIID}}
\newcommand{\gammaVint}{\eqref{eq:gamma-V-int}}
\newcommand{\figmfimaprx}{\ref{fig:mfim-aprx}}

\renewcommand{\emph}[1]{\textit{#1}}



\begin{document}
\title{Integrals of motion as slow modes in dissipative many-body operator dynamics}
\author{Tian-Hua Yang}
\author{Dmitry A. Abanin} \email{dabanin@princeton.edu}
\affiliation{Department of Physics, Princeton University, Princeton, NJ 08544, USA}

\begin{abstract}

We consider Lindbladian operator dynamics in many-body quantum systems with one or more integrals of motion (IOM), subject to weak local dissipation. We demonstrate that IOMs with small support become slow modes of these dynamics, in the sense that their Frobenius norms decay more slowly compared to generic operators. As a result, the eigenoperators of such Lindbladians with slowest decay rates have a large overlap with the IOMs of the underlying Hamiltonian. We demonstrate this correspondence between slow modes and IOMs numerically for a number of many-body models, and further corroborate it with perturbative arguments. These results open up a new method for the identification of IOMs, and provide insights into the dissipative many-body dynamics. 

\end{abstract}

\maketitle

\section{Introduction}

Recent experimental advances have made it possible to engineer synthetic quantum many-body systems and study far-from-equilibrium real-time dynamics in them~\cite{jakschColdAtomHubbard2005,blochManybodyPhysicsUltracold2008,ciracColdAtomSimulation2010,weimerRydbergQuantumSimulator2010,blochQuantumSimulationsUltracold2012,zoharColdAtomQuantumSimulator2013a,xuEmulatingManyBodyLocalization2018,browaeysManybodyPhysicsIndividually2020,schollQuantumSimulation2D2021a,daleyPracticalQuantumAdvantage2022,rosenbergDynamicsMagnetizationInfinite2024,andersenThermalizationCriticalityAnalogue2025,miNoiseresilientEdgeModes2022}. These quantum systems are inevitably subject to weak external noise~\cite{breuerTheoryOpenQuantum2009}. The word ``noise'' often bears a negative connotation. For example, it is well known that noise can prevent achieving quantum advantage~\cite{aharonovLimitationsNoisyReversible1996,harrowRobustnessQuantumGates2003,gaoEfficientClassicalSimulation2018,nohEfficientClassicalSimulation2020,deshpandeTightBoundsConvergence2022}, in the sense that broad classes of noisy quantum circuits can be classically simulated in polynomial time. In quantum simulation experiments, noise  affects the physical observables that we wish to measure~\cite{mcardleErrorMitigatedDigitalQuantum2019,vovroshSimpleMitigationGlobal2021,bennewitzNeuralErrorMitigation2022,caiQuantumErrorMitigation2023}. Therefore, it is of interest to study noisy dynamics, in order to understand how a particular kind of noise affects the outcome of an experiment. 

Intriguingly, recent works brought forth a new layer of motivation for studying noisy dynamical processes. It was discovered that (weak) dissipation can, somewhat counterintuitively, be helpful in the sense that its presence allows one to identify non-trivial features of unitary dynamical processes. For example, in a recent experiment~\cite{miNoiseresilientEdgeModes2022} which simulated the Floquet transverse-field Ising model (TFIM), it was found that one can identify the Majorana zero mode by finding the Pauli strings that decay slowly under noise. In a different direction~\cite{moriLiouvilliangapAnalysisOpen2024a,jacobySpectralGapsLocal2025,yoshimuraTheoryIrreversibilityQuantum2025,duhRuellePollicottResonancesDiffusive2025}, it was found that the Lindbladian spectrum of a chaotic local unitary quantum channel with infinitesimal local noise exhibits a non-trivial gap, argued to be the inverse of the thermalization time scale of the unitary dynamics. In both developments, the noisy operator dynamics has helped to discover or define a feature of the unitary dynamics.

Inspired by these developments, here we investigate the operator dynamics of a many-body Hamiltonian system subject to noise. We adopt the Lindbladian master equation as a description for such dynamics~\cite{goriniCompletelyPositiveDynamical1976,breuerTheoryOpenQuantum2009,rivasvargasOpenQuantumSystems2012,nathanUniversalLindbladEquation2020a,manzanoShortIntroductionLindblad2020a,ikeuchiErrorBoundsUniversal2025a}, and assume that the jump operators are local. This assumption is justified provided noise at different sites is uncorrelated, and a Markovian approximation can be applied~\cite{shiraishiQuantumMasterEquation2024b}. This setting has been considered in a series of works, including Ref.~\cite{schusterOperatorGrowthOpen2023a} which studied the universality classes of noisy operator dynamics, and Refs.~\cite{wangHierarchyRelaxationTimescales2020,hartmannFateDissipativeHierarchy2024a,macieszczakTheoryMetastabilityOpen2016,macieszczakTheoryClassicalMetastability2021,macieszczakOperationalApproachMetastability2021}, which studied the structure of the spectra of such Lindbladians and phenomena related to these structures.
 
Here, we focus on the limit of weak noise. While previous studies with similar settings have mostly considered chaotic quantum channels~\cite{schusterOperatorGrowthOpen2023a,moriLiouvilliangapAnalysisOpen2024a,jacobySpectralGapsLocal2025}, here we investigate the role of the energy and other integrals of motion (IOM) in the Hamiltonian dynamics. An important feature of such weak, local noise is that the decay rate of the Frobenius norm, or Loschmidt echo, of an operator is roughly proportional to the size of its support. Combined with unitary dynamics, this gives rise to a qualitative distinction in the growth and decay behavior of the IOMs compared to generic, non-conserved operators.
Under generic chaotic dynamics, operators undergo operator growth~\cite{nahumOperatorSpreadingRandom2018b,chenOperatorScramblingQuantum2018a,khemaniOperatorSpreadingEmergence2018c,parkerUniversalOperatorGrowth2019a,schusterOperatorGrowthOpen2023a}.
Thus, the sizes of their supports increase with time, leading to faster noise-induced decay~\cite{schusterOperatorGrowthOpen2023a}. In contrast, the IOMs are conserved under unitary dynamics and their support does not grow. Therefore, IOMs with small support would decay more slowly than a generic growing operator~\cite{miNoiseresilientEdgeModes2022,jacobySpectralGapsLocal2025}.

Exploiting this property, we demonstrate that all IOMs with support on a few lattice sites (see below for an exact definition), which we will call ``small-sized IOMs'', emerge as slow modes in weakly dissipative operator dynamics. This includes both exact IOMs and approximate IOMs, which are operators whose commutator with the Hamiltonian has a small norm. Our central methodology is exact diagonalization of the Lindbladian superoperator~\cite{barnettSpectralDecompositionLindblad2000,mingantiSpectralTheoryLiouvillians2018,nakagawaExactLiouvillianSpectrum2021,hagaLiouvillianSkinEffect2021b,costaSpectralSteadystateProperties2023}. We demonstrate that for a wide range of physically relevant models, the first few eigenoperators corresponding to the slowest decay rates, which we will call ``low-lying eigenoperators'', would have a perfect correspondence with the first few small-sized IOMs.

In addition to bringing new understanding of noisy many-body dynamics, these results allow us to systematically identify the local IOMs for a Hamiltonian, which is a problem of intrinsic value. This can be done on a classical computer by diagonalizing a weak-dissipation Lindbladian. Moreover, such Lindbladians can be implemented on state-of-the-art quantum simulators, pointing to a way of determining local IOMs on such quantum platforms.

This paper is organized as follows. In Section~\ref{sec:slow-mode-opdyn}, we set up the basic formalism of our work, and provide a qualitative picture of why IOMs emerge as slow modes of the Lindbladian. In Section~\ref{sec:slow-modes-lindblad}, we present our main results: the Lindbladian spectra of several many-body models, including the TFIM, the mixed-field Ising model (MFIM), and Heisenberg-type spin chains. We analyze the structure of the spectra, and demonstrate that the low-lying eigenoperators coincide with the small-sized IOMs. In Section~\ref{sec:discuss}, we develop a perturbative theory to justify the eigenoperator-IOM correspondence, provide the condition for exact and approximate IOMs to become eigenoperators, and discuss the implications of these conditions. Finally, in Section~\ref{sec:conclude} we provide conclusions and point to several interesting future directions.

\section{Slow modes in operator dynamics}
\label{sec:slow-mode-opdyn}

\begin{figure*}[!htbp]

\centering
\includegraphics[width=\textwidth]{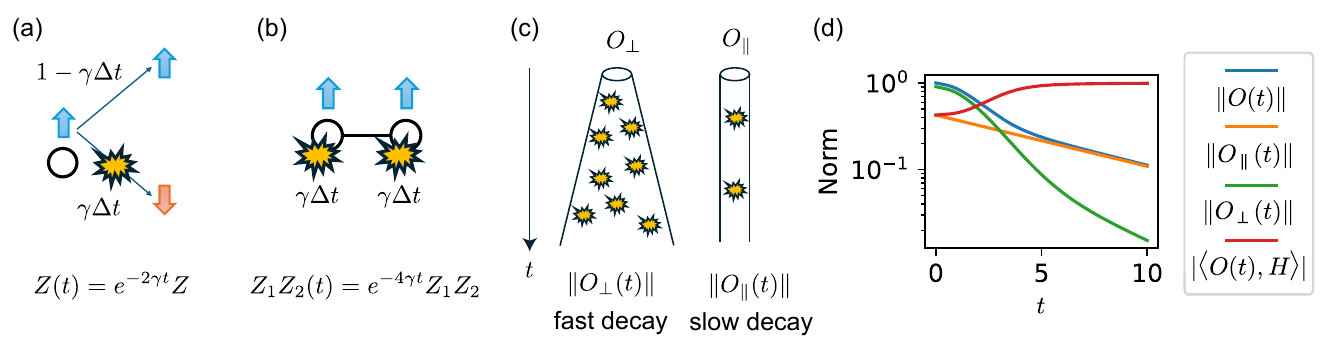}
\caption{(a) A schematic illustration of the decay of a local operator under noise. A bit flip noise results in an exponential decay in the operator $Z$. (b) For an operator that acts on multiple sites, a noise occurring on every single site can change the value of that operator, leading to a faster decay of that operator. (c) A generic operator (e.g. one that is orthogonal to all IOMs) $O_\perp$ undergoes operator growth, resulting in fast decay at late times; in contrast, an operator $O_\parallel$ that is an IOM will not grow, hence decaying more slowly. (d) A generic operator $O$ can be decomposed into its projection onto IOMs, $O_\parallel$, and the remainder $O_\perp$. Upon evolution, $O_\perp$ becomes much smaller in norm compared to $O_\parallel$, and $O(t)$ converges to $O_\parallel(t)$. This is demonstrated with a MFIM with $J=1$, $h_x=1.1$, $h_z=0.7$, where $O=\sum_i Z_i$, $O_\parallel$ is the projection of $O$ onto the total energy operator $H$, and $O_\perp = O- O_\parallel$. $\langle O(t),H\rangle = \frac{\mathrm{tr}(O(t) H)}{\sqrt{\mathrm{tr}O(t)^2 \mathrm{tr}H^2}}$ measures the overlap between $O(t)$ and $H$, which is shown to converge to $1$ due to $H$ being a slow mode.}
\label{fig:overview}
\end{figure*}

To set the stage, we consider a many-body system described by the Lindbladian master equation. We denote system's Hamiltonian by $H$ and jump operators by $\{\sqrt{\gamma_j} L_j\}$, where parameters $\gamma_j$ characterize the strength of dissipation. Generally, below we will assume that these parameters are small. The Lindbladian superoperator in the Heisenberg picture can be written as
\begin{multline}
\frac{\mathrm dO}{\mathrm dt} = \luvlD [O] \\
=i[H,O] + \sum_j \gamma_j \left(L_j^\dagger OL_j - \frac{1}{2}\{L_j^\dagger L_j,O\}\right). \label{eq:LvlDagger}
\end{multline}
We write $\luvlD$ here, since this superoperator is the Hermitian conjugate of the usual Lindbladian superoperator $\luvl$ acting on density matrices. We also write $\mathcal L^\dagger[O]=i\mathcal H[O] + \mathcal D[O]$, with $\mathcal H$ and $\mathcal D$ denoting the Hamiltonian and dissipation parts of the Lindbladian, respectively.

One important aspect of Eq.~\eqref{eq:LvlDagger} is that when the jump operators $L_j$ are local, the decay rate of an operator is roughly proportional to the size of its support. To see this, one can first rewrite the dissipative term as follows:
\begin{equation}
\mathcal D[O] = \sum_j \frac{\gamma_j}{2} \left([L_j^\dagger,O]L_j + L_j^\dagger [O,L_j]\right).
\end{equation}
Since operators that act on different sites commute with each other, each term in the parentheses vanishes unless $O$ acts non-trivially on the site(s) that $L_j$ acts on. Therefore, the dissipation term $\mathcal D[O]$ will be roughly proportional to the number of sites where $O$ acts non-trivially. To refine this argument, consider the depolarizing noise channel, which will be the default choice of noise in this paper unless otherwise stated. The depolarization noise on a qubit chain of length $N$ consists of $3N$ jump operators, equal to the Pauli operators $X_i$, $Y_i$, and $Z_i$ on each site, and all the dissipation rates are set to be the same, $\gamma_j=\gamma/4$. In this case, for any Pauli string $P$, $\mathcal D[P]=-\gamma  S(P) P$, where $S(P)$ is the size of the Pauli string, defined by the number of non-identity operators in the Pauli string. Any operator $O$ can be decomposed as a sum of Pauli strings,
\begin{equation}
O = \sum_P c_P(O) P.
\end{equation}
If $O$ is a Hermitian operator, the coefficients $c_P(O)$ would be real. Let the inner product between two operators be $(A|B) = \frac{1}{D}\mathrm{tr}(A^\dagger B)$, where $D$ is the Hilbert space dimension. Then the Frobenius norm of $O$ is
\begin{equation}
\|O\|_F=\sqrt{(O|O)} = \sqrt{\sum_P |c_P(O)|^2}.
\end{equation}
We will exclusively use the Frobenius norm in this work, hence we would drop the $F$ subscript hereafter. All Pauli strings have unit norm with this definition. The decay rate of the norm $\|O(t)\|$ under depolarizing noise is given by:
\begin{equation}
\frac{\mathrm d \|O\|}{\mathrm dt} = \frac{1}{\|O\|} \re \left( O\middle|\frac{\mathrm dO}{\mathrm dt} \right) = - \gamma S(O) \|O\|, \label{eq:O-norm-decay}
\end{equation}
where the size of an operator $S(O)$ is defined as the weighted average of the sizes of the Pauli strings contained in $O$,
\begin{equation}
S(O)=\frac{\sum_P |c_P(O)|^2  S(P)}{\sum_P |c_P(O)|^2}. \label{eq:def-op-size}
\end{equation}
Thus, under a local noise, operators of a small size will decay more slowly than large-sized ones, as illustrated in Fig.~\ref{fig:overview}(a,b). Notably, the concept of small-sized operators is more general than that of local operators. Local operators, as well as extensive sums of local operators, are clearly small-sized; however, small-sized operators are not necessarily local. For example, bi-local operators like $\sum_{i<j} Z_i Z_j$ are also small-sized. In particular, a product of two small-sized operators remains small-sized---unlike local operators, whose products need not remain local.

It is believed, and has been verified in broad classes of models~\cite{nahumOperatorSpreadingRandom2018b,chenOperatorScramblingQuantum2018a,khemaniOperatorSpreadingEmergence2018c,parkerUniversalOperatorGrowth2019a,schusterOperatorGrowthOpen2023a}, that under chaotic unitary dynamics, operator size grows with time, typically ballistically,  $S(O(t))\sim v_B t$, with some velocity $v_B$. According to Eq.~\eqref{eq:O-norm-decay}, this implies that the operator's norm decays superexponentially as $\sim e^{-\frac{1}{2}\gamma v_B t^2}$. We note that this observation played a central role in previous works on noisy operator dynamics~\cite{schusterOperatorGrowthOpen2023a,moriLiouvilliangapAnalysisOpen2024a,schusterPolynomialtimeClassicalAlgorithm2024,jacobySpectralGapsLocal2025}. In contrast, when an IOM is present, the corresponding operator commutes with the Hamiltonian and therefore does not grow under unitary dynamics. As a result, the IOM operator's norm decay as $e^{-\gamma S t}$, where $S$ is the size of the IOM. Comparing the two cases,  we find that the noise-induced decay of the IOM is parametrically slower than that of the growing operator.

This argument suggests the following picture of  noisy operator dynamics in a system with IOMs illustrated in Fig.~\ref{fig:overview}(c,d). We decompose a generic operator $O$ as $O=O_\parallel + O_\perp$, where $O_\parallel$ denotes the projection of $O$ onto the space of IOMs, and $O_\perp$ denotes the rest. We expect that $O_\perp$'s norm decays much faster than $O_\parallel$ at late times, as illustrated in Fig.~\ref{fig:overview}(c,d). After a certain period of time, $O_\perp (t)$ will have decayed to a negligible value, leading to $O(t)\approx O_\parallel(t)$. It is in this sense that we call IOMs $O_\parallel$ the slow modes. This picture will be refined and analyzed more rigorously in Section~\ref{sec:discuss}.

\section{Slow modes in the Lindbladian spectrum} \label{sec:slow-modes-lindblad}

\begin{figure*}[!htbp]

\centering
\includegraphics[width=\textwidth]{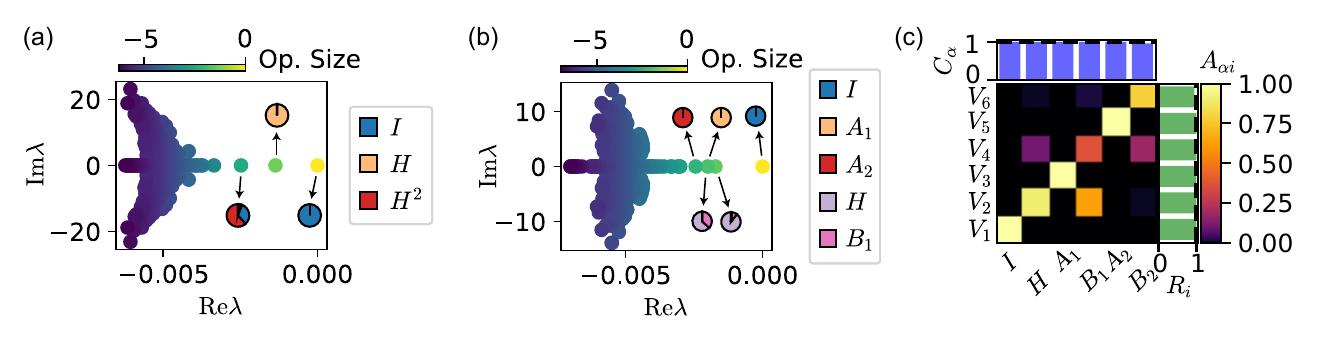}
\caption{(a) The Lindbladian spectrum of the MFIM with depolarizing noise, with parameters $J=1$, $h_x=1.1$, $h_z=0.7$, and $\gamma=10^{-3}$, on a $L=7$ PBC chain, for the zero-momentum sector. The decay rates $|\re \lambda|$ of the eigenoperators are approximately proportional to the operator size defined in Eq.~\eqref{eq:def-op-size}, shown by the color of the dots. Slowly decaying eigenmodes are discrete and well-separated from the bulk of the spectrum. The first three low-lying eigenmodes are well-approximated by linear combinations of $I$, $H$, and $H^2$, the first three IOMs. (b) The Lindbladian spectrum of the TFIM, with $J=1$, $h=0.5$, and $\gamma=10^{-3}$, on a $L=7$ chain for the zero-momentum sector. The first five eigenmodes are shown to be linear combinations of the first five IOMs. (c) The overlap between the first few low-lying eigenoperators $V_i$ with the known IOMs, for a TFIM with parameters as in (b), of size $L=9$. The colormap shows the overlap $A_{\alpha i}=|(V_i|Q_\alpha)|^2$ between the eigenoperator $V_i$ and IOM $Q_\alpha$, assuming both are normalized. The green bars $R_i$ are the ``row sums'' of $A_{\alpha i}$. If the IOMs $Q_\alpha$ are not orthogonal to each other, we first orthonormalize the set $\{Q_\alpha\}$ into $\{\tilde Q_{\alpha}\}$, and take $R_i = \sum_\alpha |(V_i | \tilde{Q}_\alpha)|^2$. If $R_i=1$, the eigenoperator $V_i$ is a linear combination of the $Q_\alpha$'s, without admixture of other, non-conserved operators. The blue bars $C_\alpha$ are the ``column sums'' of $A_{\alpha i}$ in the same fashion.}
\label{fig:lindspec-exact}

\end{figure*}

The argument given above suggests that there is a close connection between small-sized IOMs of a system's Hamiltonian and the slow modes in the dissipative operator dynamics. In this Section, we substantiate it by studying the exact Lindbladian spectra of several models, including chaotic and integrable ones~\footnote{By ``integrable'', we mean that the Hamiltonian itself is integrable, and not the full Lindbladian.}.

From now on, we will refer to small-sized IOMs simply as IOMs, unless otherwise stated. We claim that the low-lying eigenoperators of the Lindbladian superoperator $\mathcal L^\dagger$ correspond to the IOMs of the underlying Hamiltonian.
By low-lying eigenoperators, we mean eigenoperators satisfying $\mathcal L^\dagger[V]=\lambda V$ that have a slow decay rate. More precisely, we require that the decay rate $-\re\lambda = O(1)\times \gamma$, where factor $O(1)$ is a constant which does not scale with the system size $L$. We will also refer to the low-lying eigenoperators as the first few eigenoperators. Similarly, IOMs are understood as operators that commute with the Hamiltonian, whose sizes, as defined in Eq.~\eqref{eq:def-op-size}, are $O(1)$.

We start by offering a concise justification of this correspondence, with more rigorous arguments deferred to the next section and the supplementary material~\footnote{See supplementary online material.}.
On one hand, if $Q$ is an IOM of the system, satisfying $[H,Q]=0$, then, at first order in $\gamma$, we can expect that there is an eigenoperator $V\approx Q$ with an eigenvalue $\lambda \approx \frac{(Q|\mathcal D[Q])}{\|Q\|^2}$. As we have chosen the dissipation to be depolarizing noise, this reduces to $\lambda=-\gamma S(Q)$. Therefore, there is a low-lying eigenoperator $V$ that is close to $Q$, provided that $\gamma$ is small.
Conversely, if $Q$ is an operator corresponding to a low-lying eigenvalue $\lambda$, we can show that $S(Q)= \frac{|\re \lambda|}{\gamma}$, which is an $O(1)$ quantity; furthermore, under mild conditions~\cite{Note2}, $\|[H,Q]\|=O(1)\times \gamma$. Therefore, $Q$ is at least an approximate IOM, and would be an exact one if it remains a slow mode in the limit $\gamma \to 0$.

Therefore, by looking at the spectrum of the Lindbladian $\mathcal L^\dagger$, it is possible in principle to identify all IOMs, and this connection is illustrated with several examples below. All results in this Section will assume that the noise channel is the depolarizing channel, and that we imposed periodic boundary condition (PBC) and examine the zero-momentum sector. For technical discussions about different noise channels, non-zero-momentum sectors, and finite-size effects, refer to the supplementary material~\cite{Note2}.

\subsection{Mixed-field Ising model (MFIM)}\label{SecIIIA}

As the first example, we consider the MFIM, given by the Hamiltonian
\begin{equation}
H_\text{MFIM} = \sum_i J Z_i Z_{i+1} + h_z Z_i + h_x X_{i}.
\end{equation}
This model is known to be thermalizing~\cite{banulsStrongWeakThermalization2011,kimBallisticSpreadingEntanglement2013,kimTestingWhetherAll2014b} and to have no local IOMs apart from the energy itself~\cite{chibaProofAbsenceLocal2024} provided that none of $(J,h_z,h_x)$ vanish. Therefore, the first few IOMs would be the identity operator $I$, the energy $H$, and---considering products of small-sized operators are still small-sized---the first few powers of $H$. Hence, we expect that the first few eigenoperators with low-lying eigenvalues will be $I$, $H$, $H^2$, etc., or linear combinations of them. In Fig.~\ref{fig:lindspec-exact}(a), this is confirmed: the first three Lindblad eigenoperators are superpositions of these three IOMs, up to small corrections.

\subsection{Transverse-field Ising model (TFIM)}\label{SecIIIB}

\begin{figure*}[!htbp]

\centering
\includegraphics[width=\textwidth]{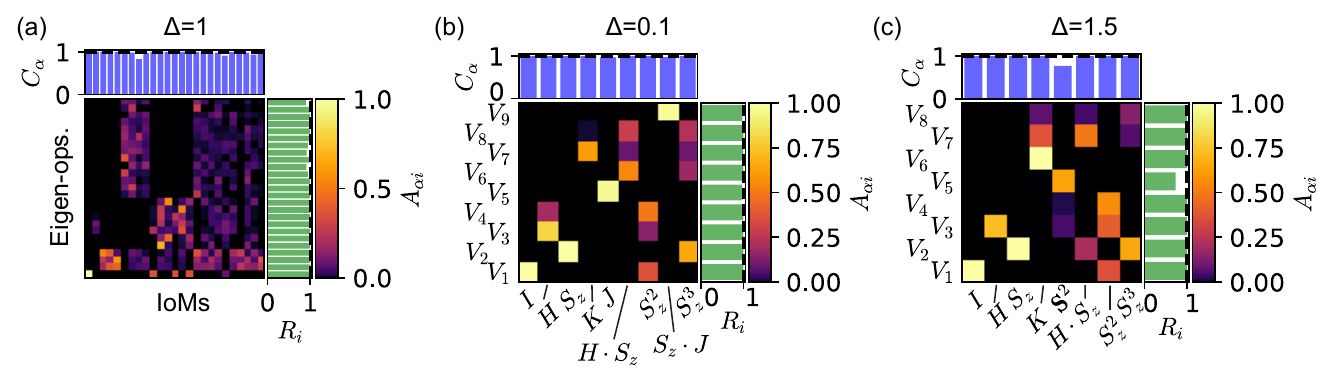}
\caption{Analysis of the low-lying eigenoperators for Heisenberg XXX and XXZ spin chains of length $L=9$. For all models, the noise is the depolarizing channel with $\gamma=10^{-3}$. The anisotropy parameter is (a) $\Delta=1$ (XXX model), (b) $\Delta=0.1$, and (c) $\Delta=1.5$. Plot legends are the same as in Fig.~\ref{fig:lindspec-exact}(c). The decomposition of the eigenoperators indicates that they are well-approximated by linear combinations of exact and approximate IOMs. See supplementary material~\cite{Note2} for more details.}
\label{fig:lindspec-hsb}

\end{figure*}

Further, we study the TFIM, with the Hamiltonian
\begin{equation}
H_\text{TFIM} = \sum_i J Z_i Z_{i+1} + h X_{i}.
\end{equation}
This model is equivalent to the Kitaev chain under a Jordan-Wigner transformation~\cite{suzukiQuantumIsingPhases2013,pfeutyOnedimensionalIsingModel1970,kitaevUnpairedMajoranaFermions2001}, and is therefore integrable. With open boundary conditions, the Kitaev chain exhibits topological Majorana edge modes~\cite{kitaevUnpairedMajoranaFermions2001,nayakNonAbelianAnyonsTopological2008,miNoiseresilientEdgeModes2022}. Under PBC, which we consider here, the following Majorana bilinear operators commute with the Hamiltonian~\cite{Note2}:
\begin{align}
A_m = &YX^{m-1}Z-ZX^{m-1}Y\hphantom{+XX}(m \geq 1), \\
B_m = & JZX^{m}Z-hZX^{m-1}Z-hYX^{m-1}Y  \notag \\&+JYX^{m-2}Y\hphantom{X^{m-1}ZXXX}(m \geq 2),\\
B_1 = & JZXZ-hZZ-hYY-JX.
\end{align}
Here, a shorthand for Pauli strings is used where, for example, $YX^m Z$ stands for $\sum_i Y_i X_{i+1} \dots X_{i+m} Z_{i+m+1}$. Numerical diagonalization of the model, with results shown in Fig.~\ref{fig:lindspec-exact}(b,c), confirms that the first few eigenoperators are, to a good approximation, linear combinations of the first few Majorana bilinear IOMs, including the energy $H$. 

\subsection{Heisenberg spin chains}\label{SecIIIC}

The XYZ Heisenberg spin chains~\cite{xiangThermodynamicsQuantumHeisenberg1998,klmperThermodynamicsAnisotropicSpin11993,affleckCriticalTheoryQuantum1987,bonnerGeneralizedHeisenbergQuantum1987} are a family of Hamiltonians given by
\begin{equation}\label{eq:H_XYZ}
H_\text{XYZ} = \sum_i J_x X_i X_{i+1} + J_y Y_i Y_{i+1} + J_z Z_i Z_{i+1}.
\end{equation}
This model exhibits different behavior for different values of couplings $J_x$, $J_y$, and $J_z$. The case where $J_x=J_y=J_z=J$ is called the XXX model, or the isotropic Heisenberg model, and is characterized by $SU(2)$ rotation symmetry. The choice $J_x=J_y\neq J_z$ corresponds to an XXZ model. In this case, an anisotropy parameter $\Delta=J_z/J_x$ is often defined. If none of $J_x$ ,$J_y$, and $J_z$ are equal to each other, the model (\ref{eq:H_XYZ}) is called the XYZ model. In all instances, the model is known to be integrable~\cite{betheZurTheorieMetalle1931,baxterOneDimensionalAnisotropicHeisenberg1971,baxterEightVertexModelLattice1971,yangGroundStateEnergyHeisenbergIsing1966}. The first few IOMs are~\cite{grabowskiQUANTUMINTEGRALSMOTION1994}:
\begin{itemize}
    \item Total spin $S_x = \sum_i X_i$ and $S_y = \sum_i Y_i$ (for the XXX model),
    \item Total spin $S_z=\sum_i Z_i$ (XXX and XXZ models),
    \item Energy $H$ (all models),
    \item $K = \sum_i \mathbf S_i \cdot (\mathbf {\tilde S}_{i+1} \times \mathbf S_{i+2})$, where $\mathbf S_i = X_i \hat x + Y_i \hat y+Z_i \hat z$, and $\mathbf{\tilde S_i} = J_x^{-1} X_i\hat x+J_y^{-1} Y_i \hat y+J_z^{-1} Z_i \hat z$ (all models).
\end{itemize}

\begin{figure*}[!htbp]

\centering
\includegraphics[width=\textwidth]{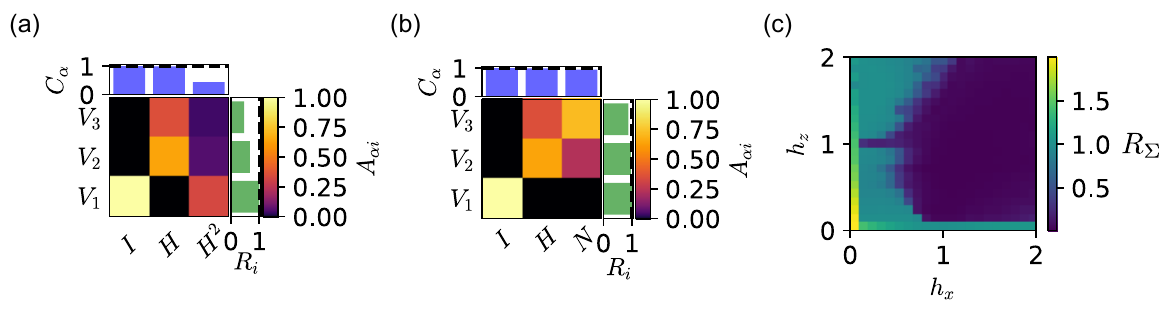}
\caption{Illustration of finding approximate IOMs in the MFIM. (a) The first three Lindbladian eigenoperators overlaps with $I$, $H$, and $H^2$, are shown, for model parameters $J=1$, $h_z=h_x=0.4$, chain length $L=8$, subject to depolarizing noise with $\gamma=10^{-3}$. Plot legends are the same as in Fig.~\ref{fig:lindspec-exact}(c). The second and third eigenoperators are not linear combinations of known IOMs. Note that the apparent overlap between $H^2$ and $V_1$ is due to the fact that $H^2$ has overlap with $I$. (b) Similar to (a), but we take into account that this system has an approximately conserved ``dressed particle number operator'' $N$ (see Section V of Ref.~\cite{wurtzEmergentConservationLaws2020} or the supplementary materials~\cite{Note2}). We find that the second and third eigenoperators are well-represented by linear combinations of $H$ and $N$. (c) A ``chaoticity phase diagram'' of the MFIM, obtained by plotting $R_\Sigma=\sum_{i=1}^3(1-R_i)$ for models with different $h_x$ and $h_z$ while fixing $J=1$. $R_\Sigma$ is expected to vanish if the model is fully chaotic. We observe that the model is fully chaotic for $h_x>1$ and $h_z\neq0$. On the other hand, for large regions where $0<h_x<1$ and $h_z\neq 0$, the system has non-$H$-power slow modes, despite the absence of other exact local IOMs.}
\label{fig:mfim-aprx}

\end{figure*}

For the XXX model, these known IOMs (together with their products) account for the first 25 eigenoperators in the spectrum, as is shown in Fig.~\ref{fig:lindspec-hsb}(a). However, for the XXZ model, we notice that the IOMs listed above are not sufficient to account for all slow modes. By examining the eigenoperators that do not overlap with the listed IOMs, we uncover several approximate IOMs in different ranges of $\Delta$. When $\Delta$ is small, the system resembles an XX model, which is equivalent to a free fermion chain and hence possesses many conserved quantities of fermion-bilinear operators. We find that the first one, the particle current operator $J=XY-YX$, appears as a slow mode for a $\Delta=0.1$ XXZ model, as shown in Fig.~\ref{fig:lindspec-hsb}(b). For a model with $\Delta=1.5$, as presented in Fig.~\ref{fig:lindspec-hsb}(c) , we find that the first ``additional'' IOM is approximately equal to $S_x^2+S_y^2$. Considering $S_z$ is exactly conserved, this is equivalent to saying that the operator $\mathbf S^2 = S_x^2+S_y^2+S_z^2$, the Casimir operator associated with the global $SU(2)$ symmetry in the isotropic case, is approximately conserved for $\Delta\approx 1$. A more detailed discussion of our findings for Heisenberg spin chains is presented in the supplementary material~\cite{Note2}.

\subsection{Probing approximate conservation laws}\label{SecIIID}

As we have just demonstrated, the content of low-lying eigenoperators can reveal exact and approximate IOMs alike. Hence, it can be used as a tool to determine the presence (absence) of approximate IOMs. For example, consider the MFIM. The model is integrable if $h_z=0$, where it reduces to the TFIM, and becomes classical if $h_x=0$. Otherwise, it is believed to be non-integrable. However, this does not mean that the system would always behave like a fully chaotic model. In fact, approximate IOMs and prethermalization behavior are known to exist in certain parameter ranges~\cite{wurtzEmergentConservationLaws2020,colluraDiscreteTimeCrystallineResponse2022}.

Generally, a many-body model is expected to be chaotic if it has no other local IOM apart from the Hamiltonian $H$ itself. Our analysis above suggests a somewhat refined measure of chaoticity, based on the low-lying Lindbladian spectra at weak dissipation.
Assuming that $H$ is the only physically relevant conservation law, we would expect that the only small-sized IOMs are $I$, $H$, and the first few powers of $H$. Therefore, in a fully chaotic model, the first few eigenoperators should be a linear combination of these powers of $H$. Otherwise, this model would have additional (approximate) conservation laws. We tested this for the first three eigenoperators of a MFIM with $J=1$ and $h_x=h_z=0.4$. Indeed, we found that the first three eigenoperators are not fully accounted for by inserting powers of $H$. Instead, they contain an approximate conservation law, which is identified with the renormalized particle number operator proposed in ~\cite{wurtzEmergentConservationLaws2020}. This is demonstrated in Fig.~\ref{fig:mfim-aprx}(a,b).

Generalizing this example, we propose a measure of how chaotic a model is: we take the first few low-lying eigenoperators in the Lindbladian spectrum, and project them onto the space spanned by first few powers of the Hamiltonian. If these two sets of operators are the same up to a linear combination, this model is considered fully chaotic. The difference between the two sets is a measure of how non-chaotic the model is. This measure is somewhat more general than the traditional definition of chaoticity, as it accounts not only for the exact IOMs, but also approximate symmetries. For example, if the system displays a prethermalization plateau protected by an approximate symmetry~\cite{abaninRigorousTheoryManyBody2017a} up to long times, but thermalizes eventually, it would be treated as chaotic by the traditional measure, while our measure would classify it as not fully chaotic.

Using this measure, we draw a ``chaoticity phase diagram'' of the MFIM in Fig~\ref{fig:mfim-aprx}(c). We see that the two coordinate axes corresponding to vanishing $h_z$ or $h_x$ are indeed non-chaotic, while the parts where both $h_x$ and $h_z$ are large are chaotic. However, we found that large parts of the diagram where both $h_x$ and $h_z$ are non-vanishing are also not fully chaotic: approximate conservation laws exist for a wide range of parameters where the transverse field $h_x$ is weak. This shows that approximate IOMs are present in much of the MFIM phase diagram; or, put in other words, the classical Ising model does not easily become fully quantum chaotic upon the addition of a weak transverse field. A more in-depth examination of this phase diagram would be an interesting topic for further studies.

\begin{figure*}[!htbp]

\centering
\includegraphics[width=\textwidth]{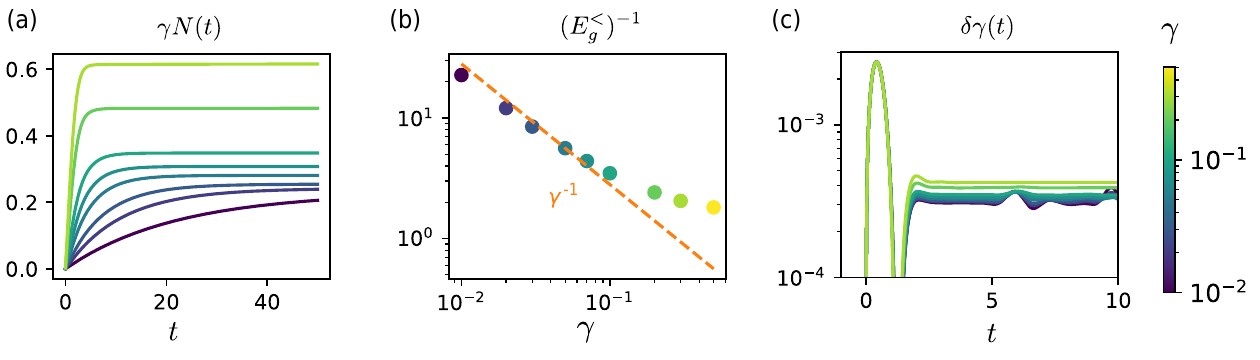}
\caption{Illustration of how an approximate IOM converges to an eigenoperator. Numerical simulation is performed for a MFIM, with $J=1$, $h_x=2$, and $h_z=0.05$, on a chain with $L=9$. Since $h_z$ is small, this model can be considered a perturbed TFIM. The approximate IOM is taken to be the Majorana bilinear operator $A_1$. Noise channel is taken to be depolarizing with amplitude $\gamma$. (a) Evaluation of $N(t)=\int_0^{t} \mathrm d\tau \|J_1(\tau)\| e^{\gamma_Q \tau}$, the integral in Eq.~\eqref{eq:V-perp-norm-bound} up to a finite time, with $\gamma_V$ approximated by $\gamma_Q$. We can observe that $\gamma N(t)$ converges to an $O(1)$ value. (b) $(E_g^<)^{-1}=\lim_{t\to \infty}N(t)$ as a function of $\gamma$. The curve has a $\gamma^{-1}$ behavior for small $\gamma$, tending upwards for large $\gamma$, in line with the behavior expected from Eq.~\eqref{eq:J1-late-time}. (c) $\delta\gamma(t) = \left|\frac{\mathrm d\|Q(t)\|}{\mathrm dt}\right| - \gamma_Q$, the decay rate of $\|Q(t)\|$ offset by the first-order contribution. We observe that it converges quickly, with the time scale of convergence and plateau value being insensitive to $\gamma$.}
\label{fig:approx-IOM}

\end{figure*}

\section{Perturbation theory and approximate symmetries} \label{sec:discuss}

Our discussion in Section~\ref{sec:slow-modes-lindblad} has been based on first-order perturbation theory. This is accurate when an exact IOM $Q$ with $[H,Q]=0$ also happens to be an exact eigenoperator of the full Lindbladian $\mathcal L^\dagger$. For example, $S_z$ and $H_\text{XYZ}$ are exact eigenoperators of the depolarizing channel, since they are sums of Pauli strings of the same size. Many of the previously-identified slow modes in Lindbladian systems also fall into this category~\cite{teretenkovExactDynamicsQuantum2024,teretenkovDualityOpenSystems2024,wangEmbeddingQuantumManyBody2024,garcia-garciaLindbladManybodyScars2025}. However, many IOMs, such as the energy $H$ for the Ising models, or any operator that only commutes approximately with $H$, do not possess this property. In these cases, the action of the Lindbladian can be written as follows, 
\begin{equation}
\mathcal L^\dagger[Q] = -\gamma_Q Q + \epsilon_1 J_1, \label{eq:Ldagger-on-Q-aprx}
\end{equation}
where $J_1$ is taken to be normalized and orthogonal to $Q$. Parameter $\gamma_Q$ can be identified with the decay rate of $Q$ in the first-order perturbation theory, and $\epsilon_1$ is another small parameter that characterizes the extent to which the Lindbladian can change $Q$ . In the cases where $Q$ is an exact IOM of $H$, this change is exclusively caused by noise, and $\epsilon_1$ would be of the same order as $\gamma$. However, the form of Eq.~\eqref{eq:Ldagger-on-Q-aprx} can also encompass the case where $Q$ does not commute exactly with $H$. For example, if $H=H_0+\delta H$, where $[H_0,Q]=0$ but $[\delta H,Q]\neq 0$, the action of the Lindbladian on $Q$ can also be described by Eq.~\eqref{eq:Ldagger-on-Q-aprx}, with $\epsilon_1\sim \|\delta H\|$. In the following discussions, we treat the cases where $Q$ is an exact IOM and where it is an approximate one, on an equal footing. 

Returning to the question of the validity of the first-order perturbation theory, it may seem that perturbative arguments should not be applicable in our many-body setting. Naively, the small parameter in the perturbation theory is $\frac{\epsilon_1}{E_g}$, where $E_g$ is the energy gap of the unperturbed Lindbladian. For quantum many-body systems, when we consider non-equilibrium processes, this energy gap would be the level spacing between mid-spectrum states, which is exponentially small in system size. Meanwhile, $\epsilon_1$ is usually insensitive to system size. This means that $E_g$ is likely to be smaller than $\epsilon_1$ for systems larger than a few spins, leading to a breakdown of perturbation theories.

Despite this apparent contradiction, our numerical results show that the spectrum closely follows the perturbative predictions. Why, then, does perturbation theory still work? A simplistic answer is that the relevant gap is not the microscopic level spacing, but rather the Mori gap—characteristic of weakly dissipative systems—which remains finite as the system size increases~\cite{moriLiouvilliangapAnalysisOpen2024a,jacobySpectralGapsLocal2025}. To develop a more complete understanding, we will examine the perturbative argument in detail.
Suppose that the operator $Q$ that satisfies Eq.~\eqref{eq:Ldagger-on-Q-aprx} is contained in an eigenoperator $V$ with eigenvalue $-\gamma_V$, which means that $(\mathcal L^\dagger + \gamma_V)V=0$. Let $V=Q+V_\perp$, where $V_\perp$ is orthogonal to $Q$. We can project the eigenvalue equation onto the subspaces parallel and orthogonal to $Q$, respectively, to obtain
\begin{align}
\gamma_V - \gamma _Q+ \epsilon_2^\ast (J_2|V_\perp) & = 0, \label{eq:gamma-V-gamma-Q} \\
(\mathcal L^\dagger_\perp +\gamma_V)V_\perp + \epsilon_1 J_1 &=0.\label{eq:V-perp-cons}
\end{align}
We have let $\mathcal L[Q] = -\gamma_Q Q + \epsilon_2 J_2$, where $J_2$ is also normalized and orthogonal to $Q$, and utilized $(Q|\mathcal L^\dagger[V]) = (\mathcal L[Q]|V)$. Above, $\mathcal L^\dagger_\perp$ is $\mathcal L^\dagger$ restricted to the subspace orthogonal to $Q$. $\epsilon_2$ is typically of the same magnitude as $\epsilon_1$, and we use $\epsilon$ to denote the order of magnitude of both quantities.

We can solve for $V_\perp$ from Eq.~\eqref{eq:V-perp-cons}, which yields
\begin{equation}
V_\perp = \epsilon_1 \int_0^{+\infty} e^{(\mathcal L^\dagger_\perp +\gamma_V)t} J_1\mathrm dt.
\end{equation}
Plugging this back to Eq.~\eqref{eq:gamma-V-gamma-Q}, we obtain
\begin{equation}
\gamma_V = \gamma_Q - \epsilon_1\epsilon_2^\ast \int_0^{+\infty} e^{\gamma_V t} \left(J_2\middle| e^{\mathcal L^\dagger_\perp t} J_1 \right) \mathrm dt.
\end{equation}
This is a self-consistent equation that determines $\gamma_V$. Roughly equating $e^{\mathcal L^\dagger_\perp t} J_1 \approx e^{\mathcal L^\dagger t} J_1 = J_1(t)$, which is expected to be a reasonable approximation in the case of small $\epsilon$, we have
\begin{align}
\gamma_V & = \gamma_Q - \epsilon_1\epsilon_2^\ast \int_0^{+\infty} e^{\gamma_V t} \left(J_2\middle| J_1(t) \right) \mathrm dt, \label{eq:gamma-V-int} \\ 
V_\perp & = \epsilon_1 \int_0^{+\infty} e^{\gamma_V t} J_1(t)\mathrm dt. \label{eq:V-perp-int} 
\end{align}
We are now in a position to justify this perturbative approach. The first-order perturbation theory is a good approximation if $\gamma_V \approx \gamma_Q$ and $\|V_\perp\| \ll \|Q\|$. Considering $|\gamma_V-\gamma_Q| = |\epsilon_2| |(J_2|V_\perp)| \leq |\epsilon_2| \|V_\perp\|$, the first condition is implied by the second. As a rough estimate,
\begin{equation}
\|V_\perp\| \lesssim |\epsilon_1| \int_0^{+\infty} e^{\gamma_V t} \|J_1(t)\|\mathrm dt. \label{eq:V-perp-norm-bound}
\end{equation}
This equation is central to our discussion about the viability of perturbation theory. For unitary dynamics, $\|J_1(t)\|$ remains constant throughout the evolution, which means that the integral on the right-hand side of Eq.~\eqref{eq:V-perp-norm-bound} always diverges no matter how small $\epsilon_1$ is. On the other hand, for dissipative systems, $\|J_1(t)\|$ would decay, and decay quickly as was argued in Section~\ref{sec:slow-mode-opdyn}. If we assume that $J_1(t)$ grows and decays in a fashion similar to the operator dynamics in fully chaotic systems, i.e., $S(J_1(t))$ grows ballistically as $v_B t$ when $v_B t<L$, and saturates to around $L$ at late times, we would have
\begin{equation}
\|J_1(t)\|\sim\begin{cases}
e^{-\frac{1}{2}\gamma v_B t^2}, & v_B t<L, \\
e^{-\gamma L t}, & v_B t>L.
\end{cases} \label{eq:J1-late-time}
\end{equation}
In both cases, the decay is faster than the growth $e^{\gamma_V t}$, since $\gamma_V$ is usually $O(1)\times \gamma$; therefore, Eq.~\eqref{eq:V-perp-norm-bound} will converge. Writing Eq.~\eqref{eq:V-perp-norm-bound} as
\begin{equation}
\|V_\perp\| \lesssim \frac{\epsilon_1}{E_g^<}, \label{eq:V-perp-small}
\end{equation}
we would expect $E_g^<\sim \min(\gamma L ,\sqrt{v_B \gamma})$~\cite{moriLiouvilliangapAnalysisOpen2024a,jacobySpectralGapsLocal2025}. This is demonstrated in Fig.~\ref{fig:approx-IOM}(a,b). In either case, the gap $E_g^<$ would not become exponentially small as we go to larger system sizes. In fact, if $\gamma$ is small and $L$ is large, we always have $E_g^<\gg \gamma$. Therefore, we reasonably expect the perturbation theory to hold as long as $\epsilon \lesssim \gamma \ll 1$.

Three comments are in order here. First, our result implies the existence of a family of slow modes given by hydrodynamics in the Mori limit $\lim_{\gamma\to 0}\lim_{L\to\infty}$~\cite{moriLiouvilliangapAnalysisOpen2024a}. For any Hamiltonian of the form $H=\sum_x h_x$,  where $h_x$ is the local energy density supported near site $x$, the modulated energy density operator $H_k = \sum_x e^{ikx} h_x$ would have an $O(k)$ commutator with $H$ when $k$ is small. As $L$ goes to infinity, $k$ becomes continuous, hence the family $H_k$ contains operators that, despite being purely off-diagonal in the energy eigenbasis~\footnote{$\langle m|H_k|m\rangle=0$ for any energy eigenstate $|m\rangle$ is obvious from momentum-sector selection rule.}, have non-zero but arbitrarily small commutators with $H$. Therefore, a continuum of slow modes would exist for any non-zero $\gamma$---the Lindbladian spectrum for a weakly-dissipative Hamiltonian system becomes gapless in the Mori limit, even if we project out all the exact IOMs. The gap wouldn't be restored even by projecting out the $H_k$'s, as there is a tower of higher-order hydrodynamic modes which are also approximate IOMs~\cite{delacretazBoundThermalizationDiffusive2025}.  This suggests that for Hamiltonian systems, the Mori gap must be defined in sectors that are orthogonal to all hydrodynamic modes~\cite{matthiesThermalizationHydrodynamicLongtime2024,mccullochSubexponentialDecayLocal2025}, and might not exist at all.

Second, the discussion above has assumed that there are no IOMs in the system other than $Q$, or at least $J_1$ and $J_2$ do not overlap with any other IOMs. This is usually an oversimplified assumption for Hamiltonian systems. In case this assumption does not hold, in principle, the above formulation can be modified by expanding the subspace of operators parallel to $Q$ to the subspace spanned by all relevant IOMs, including approximate ones, to ensure that operators in the perpendicular subspace grow and decay quickly enough. Practically, this may be challenging, especially in light of the existence of the hydrodynamic slow modes mentioned above. Nevertheless, we expect that one may consider this setting with the parallel space containing the most obvious and most local IOMs. While the operator dynamics in the perpendicular subspace might not be in line with the expectations for fully chaotic models, as long as some form of operator growth takes place such that $J_1(t)$ has a larger size than $Q$ at late times, the integral in Eq.~\eqref{eq:V-perp-norm-bound} will converge, justifying the applicability of perturbative approach.

Lastly, we point out that the condition for the convergence of the integral in Eq.~\eqref{eq:gamma-V-int} is usually much looser than that for the integral in Eq.~\eqref{eq:V-perp-int}. For example, as demonstrated in Fig.~\ref{fig:approx-IOM}(a), the time it takes for $V_\perp$ to converge to its stationary form becomes very long for small $\gamma$; however, as shown in Fig.~\ref{fig:approx-IOM}(c), the decay rate of $\|Q(t)\|$ stabilizes at a rate that is nearly independent of $\gamma$. This is due to the fact that the decay of $\|J_1(t)\|$ is fully due to dissipation, yet $(J_2|J_1(t))$ often decays rapidly even without dissipation. If we call the time scale at which $(J_2|J_1(t))$ decays as $E_g^>$, it is expected that the decay rate of $\|Q(t)\|$ would stabilize to $\gamma_V$ on a time scale $(E_g^>)^{-1}$, as long as $\epsilon \ll E_g^>$. Therefore, for $E_g^< < \epsilon < E_g^>$, although $Q$ is not present as an exact eigenoperator of the Lindbladian operator, it can be considered an approximate eigenoperator. Notably, $E_g^>$ remains finite as $\gamma \to 0$, meaning that this quasi-eigenoperator picture can also be used to describe weak integrability breaking in dissipationless systems. In fact, to the lowest order in the symmetry breaking parameter $\epsilon$, the eigenvalue is given by
\begin{equation}
\gamma_V = - \epsilon_1\epsilon_2^\ast \int_0^{+\infty} \left(J_2\middle| J_1(t) \right) \mathrm dt. \label{eq:gamma-V-aprx-sym}
\end{equation}
This is consistent with the decay rate of an approximate IOM calculated by Fermi's golden rule~\cite{baumgartnerHilbertSpaceDiffusion2025}, and is shown to serve as a measure of the time scale on which an approximate symmetry breaks down.

\section{Conclusions and outlook}
 \label{sec:conclude}

We have demonstrated that small-sized IOMs and approximate IOMs give rise to slowly-decaying eigenoperators in the spectrum of many-body Lindbladians. We formulated a perturbation theory to calculate the decay rates of these eigenoperators, and outlined the condition for exact and approximate IOMs to become eigenoperators. Further, by performing numerical diagonalization of the Lindbladian for several classes of spin chain models, we showed that this method can reveal the well-known IOMs, as well as uncover approximate or quasi-local IOMs.

Our work can be considered as a new formalism within the research theme of identifying symmetries using superoperators~\cite{lianConservedQuantitiesEntanglement2022a,chenConservedQuantitiesGeneralized2024,moudgalyaSymmetriesGroundStates2024}. One notable feature our setting is that the Lindbladian superoperator adopted in our work has a clear physical meaning, and can be readily implemented, either numerically, or experimentally in quantum simulation platforms. It would be interesting to compare our approach to those existing in the literature or to interpret existing methodologies under a picture of (dissipative) operator dynamics.

The analysis in this paper utilized exact diagonalization of the Lindbladian superoperator. While this methodology is restricted to small system sizes, the physical picture is applicable to arbitrary system sizes. We note that, in the future, larger systems can be analyzed in two ways. On the numerical side, it is possible to employ matrix product state algorithms to perform the diagonalization. This can be done either by employing time-evolution block-decimation~\cite{paeckelTimeevolutionMethodsMatrixproduct2019} to perform the Lindbladian time evolution to late times, or by using the density matrix renormalization group~\cite{schollwockDensitymatrixRenormalizationGroup2011a} to directly find the eigenoperators of the superoperator. This will potentially allow us to search for IOMs of large-sized quantum many-body systems on a classical computer.
 
On the other hand, our method can be applied directly in quantum simulation experiments. The current technologies allow realizing Lindbladian dynamics, where the noise can either be natural or engineered~\cite{plenioCavitylossinducedGenerationEntangled1999,krausPreparationEntangledStates2008,diehlQuantumStatesPhases2008,verstraeteQuantumComputationQuantumstate2009,miStableQuantumcorrelatedManybody2024a}. We envision an experiment where an initial state $|\psi\rangle$ is prepared, evolved to time $t$ under a Lindbladian, and shadow tomography is performed~\cite{huangPredictingManyProperties2020a,zhanLearningConservationLaws2024} to measure the expectation value of any local operator $O$ of a certain weight. Therefore, expectation values like $\langle \psi | O(t)|\psi\rangle$ are measurable for any physically preparable state $|\psi\rangle$ and any small-sized operator $O$. With adequate choices of $|\psi\rangle$, it is possible to reconstruct $O(t)$ at later times. This allows to identify the operators that survive for a long time under dissipative evolution, which are exactly the Lindbladian's slow modes, corresponding to the IOMs of the system. Detailing this methodology would be left to future work.

Apart from identifying IOMs, our findings reveal valuable insights into the structure of noisy operator dynamics. Specifically, any evolving operator can be expressed as the sum of slow modes associated with IOMs and a complementary set of fast modes. By separating these contributions, one can quantitatively describe how noise corrupts physical observables over time.

\section*{Acknowledgements}

T.-H. Yang thanks Luca V. Delacr\'etaz and J. Alexander Jacoby for insightful discussions. We thank Marko Ljubotina for collaboration on a related project. Numerical calculations in this work are performed using the QuSpin library~\cite{weinbergQuSpinPythonPackage2017,weinbergQuSpinPythonPackage2019}. The data that support the findings of this article are openly available~\footnote{\href{https://github.com/ThomasYangth/Lindblad_IOM}{https://github.com/ThomasYangth/Lindblad\_IOM}}. ChatGPT has been used to improve the manuscript. 

\nocite{schusterOperatorGrowthOpen2023a,angrisaniSimulatingQuantumCircuits2025,grabowskiQUANTUMINTEGRALSMOTION1994,ilievskiCompleteGeneralizedGibbs2015a,ilievskiQuasilocalChargesIntegrable2016a,pereiraExactlyConservedQuasilocal2014}

\clearpage
\appendix
\onecolumngrid

\section*{Supplementary materials}

\section{Properties of Lindbladian eigenoperators}

In this section, we discuss several general properties regarding eigenoperators in the Lindbladian spectrum.

First, consider we have an eigenoperator $V$, characterized by
\begin{equation}
i[H,V] + \mathcal D[V] = \lambda V.
\end{equation}
We can take the Hermitian conjugate of this. Noticing that $\mathcal D[V]^\dagger =\mathcal D[V^\dagger]$,
\begin{equation}
i[H,V^\dagger] + \mathcal D[V^\dagger] = \lambda^\ast V^\dagger.
\end{equation}
This means that if $V$ is an eigenoperator of eigenvalue $\lambda$, then $V^\dagger$ is an eigenoperator of eigenvalue $\lambda^\ast$. As straightforward corollaries, (i) the spectrum of the Lindbladian is symmetric with respect to the real axis, and (ii) eigenoperators with real eigenvalues can be taken to be Hermitian.

We can also write
\begin{equation}
i \mathrm{tr}\{V^\dagger [H,V]\} + \mathrm{tr} \{V^\dagger \mathcal D[V]\} = \lambda \mathrm{tr}(V^\dagger V).
\end{equation}
We now focus only on the case where $\lambda$ is real. In this case, $V$ is Hermitian, and $\mathrm{tr}\{V[H,V]\}=0$. Without loss of generality, assume that $V$ is normalized. Then, we have
\begin{equation}
\lambda = \mathrm{tr} \{V^\dagger \mathcal D[V]\} = (V|\mathcal D[V]). \label{eq:lambda-as-trace}
\end{equation}
Consequently,
\begin{equation}
i[H,V] = (V|\mathcal D[V]) V - \mathcal D[V].
\end{equation}
Therefore,
\begin{equation}
\| [H,V] \|^2 = \|\mathcal D[V]\|^2 - ( V|\mathcal D[V]) ^2.
\end{equation}
Assume that we take the depolarizing channel, with $\mathcal D = -\gamma S$, we can write this in a more illuminating form,
\begin{equation}
\|[H,V]\|^2 = \gamma^2 \left( \langle S^2 \rangle _V - \langle S\rangle _V^2\right).
\end{equation}
Here, $\langle S^n\rangle_V = (V|S^n[V])$. If we write $V = \sum_P c_P(V) P$, where $P$ are Pauli strings, then $\langle S^n\rangle_V = \sum_P |c_P(V)|^2 S(P)^n = \sum_S p_S S^n$, where $p_S=\sum_{S(P)=S} |c_P(V)|^2$. That is, $p_S$ is a weight distribution corresponding to the sizes of Pauli strings contained in $V$, and $\langle S^2\rangle_V - \langle S\rangle_V^2$ is the variance of this distribution. For reasonable distributions, we would expect $\langle S^2\rangle_V - \langle S\rangle_V^2 \sim O(1)\times \langle S\rangle_V^2$. Therefore, if $\lambda = O(1)\times \gamma$, we can conclude that $\|[H,V]\| = O(1)\times \gamma$.

\section{Impact of noise channel and noise amplitude}

The results presented in the main text have focused exclusively on the case where the noise channel is an extremely weak depolarizing noise. In this section, we discuss how much our Lindbladian spectrum is going to change if we switch to different kinds of noise.

\paragraph{Noise amplitudes.} According to the discussion in the main text, the noise amplitude $\gamma$ would not have a large impact on the exact IOMs. In fact, since $\epsilon$ for an exact IOM scales at most with $\gamma$, the efficacy of perturbation theory for exact IOMs is not sensitive to $\gamma$. Therefore, for reasonably small $\gamma$, exact IOMs are always going to be slow modes, with a decay rate that is linear in $\gamma$ to the first order.

\begin{figure}[!htbp]
    \centering
    \includegraphics[width=\textwidth]{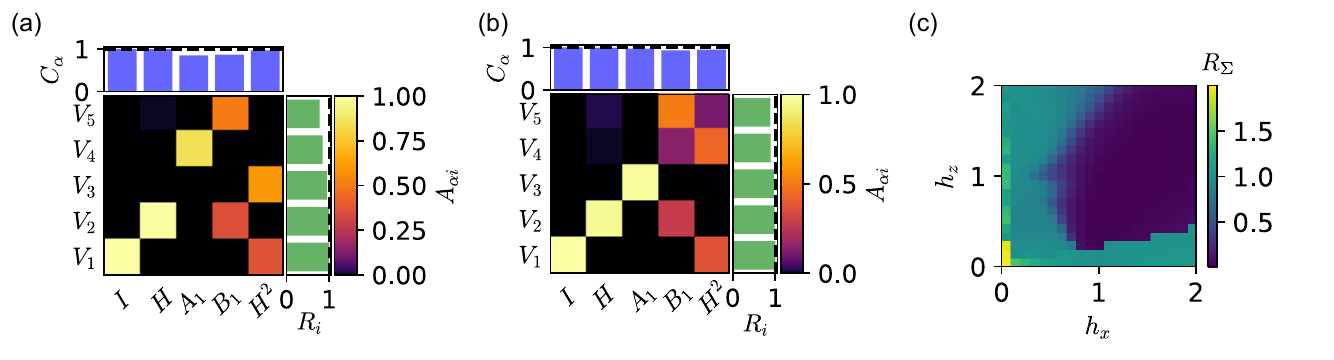}
    \caption{Lindbladian spectra of MFIM systems with stronger noise. (a) A MFIM with $J=1$, $h_x=1.5$, and $h_z=0.1$. The noise channel is the depolarizing channel with $\gamma=10^{-3}$. With $h_z$ being small, the exact IOMs of the corresponding TFIM are expected to survive as approximate IOMs. We can see that while $V_2\approx H$ and $V_3\approx H^2$, $V_4\approx A_1$, an approximate IOM. (b) Similar to (a), but with $\gamma=0.1$. We see now that $V_3\approx A_1$ and $V_4 \approx H^2+B_1$, meaning that the approximate IOMs $A_1$ and $B_1$ are favored against the exact IOM $H^2$ due to the noise being stronger. (c) The MFIM phase diagram for depolarizing noise with $\gamma=0.1$. Compared to Fig.~\figmfimaprx(c) in the main text, there is more ``non-chaotic'' region, especially near the TFIM-limit axis.}
    \label{fig:S3}
\end{figure}

The impact is more significant when we look at approximate IOMs. An approximate IOM would be a slow mode only if the noise amplitude $\gamma$ is larger than its commutator with the Hamiltonian $\epsilon$. In particular, since the decay rate of an approximate IOM roughly has the form $-\lambda =A\epsilon^2 + B\gamma$ (see Eq.~\gammaVint\ in the main text), as $\gamma$ becomes small, $\frac{|\lambda|}{\gamma} = B+A\frac{\epsilon^2}{\gamma}$ would become large. In comparison, $\frac{|\lambda|}{\gamma}$ should roughly be a constant for an exact IOM, namely the size of the IOM. Therefore, as $\gamma$ becomes smaller, approximate IOMs would gradually become less favored in the spectrum. On the flip side, as $\gamma$ becomes larger, the effect of $B$ would be more important than $A$; therefore, we expect to see more small-sized approximate IOMs appear as slow modes.

We can witness this transition in Fig.~\ref{fig:S3}. In Fig.~\ref{fig:S3}(a,b), we pick a MFIM model with weak longitudinal field, and calculate the spectrum for a weak noise and a strong noise. Due to its proximity to the TFIM, the model possesses many approximate IOMs deriving from the IOMs $A_n$ and $B_n$ of the TFIM. We see that both $A_1$ and $B_1$ exist in the spectra in Fig.~\ref{fig:S3}(a,b). Notably, in Fig.~\ref{fig:S3}(b), as the noise becomes stronger, $A_1$ gets favored over the exact IOM $H^2$, by being promoted from the 4th slowest mode to the 3rd slowest mode.

This has interesting consequence for the chaoticity measure we proposed in Section~{\SecIIID} in the main text. We have defined the chaoticity of a model as how well the first three slow modes, $\{V_1,V_2,V_3\}$, overlap with the first three exact IOMs $\{I,H,H^2\}$. If an approximate IOM gets a slower decay rate than $H^2$, this would make our measure conclude that the model is not chaotic. This can be seen in Fig.~\ref{fig:S3}(c), where we plotted a chaoticity phase diagram for the MFIM for a large noise. We see that a ``non-chaotic'' region appears near the $h_z=0$ axis, which was previously nonexistent when the noise is weak. This is exactly due to the approximate IOM $A_1$ taking over the place of $H^2$.

Physically, we can roughly say that a noisy model with noise rate $\gamma$ has similar physics with a noiseless model if we focus on a time scale $t\ll \frac{1}{\gamma}$. Taking a larger $\gamma$ means focusing on shorter time scales, and more approximate IOMs would have a larger effect on the dynamics in such time scales. Therefore, the lesson from Fig.~\ref{fig:S3}(c) can be roughly phrased as: a model that looks chaotic on a long time scale may look less chaotic at shorter times.

\begin{figure}[!htbp]
    \centering
    \includegraphics[width=\textwidth]{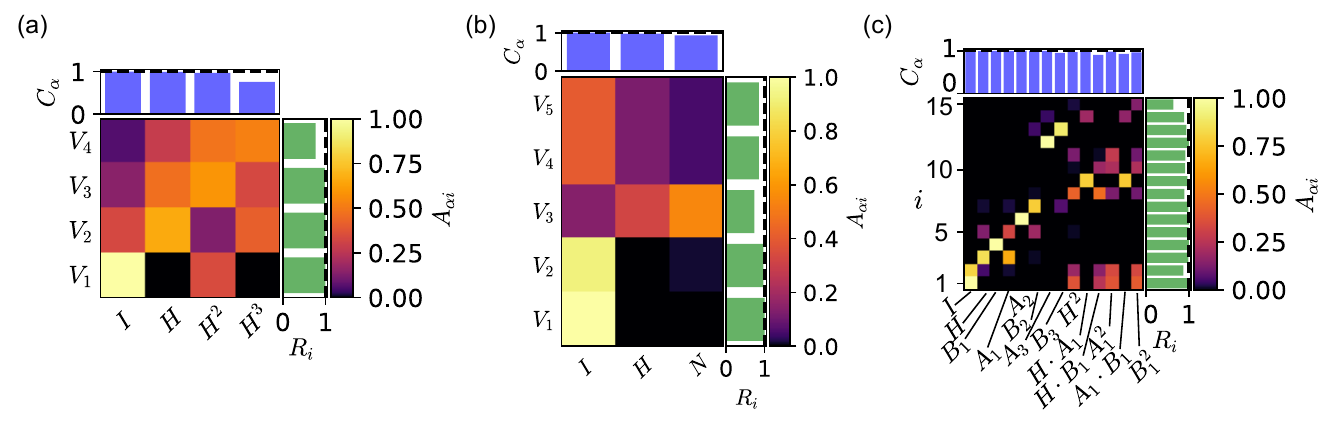}
    \caption{Lindbladian spectra for MFIM and TFIM systems with dephasing-decay noise, all of strength $\gamma=10^{-3}$. (a) MFIM with $J=1$, $h_z=0.7$, $h_x=1.1$. (b) MFIM with $J=1$, $h_z=h_x=0.4$. (c) TFIM with $J=1$, $h=0.5$.}
    \label{fig:S2}
\end{figure}

\paragraph{Noise channels.} Despite its mathematical appeal, the depolarizing noise channel is not expected to characterize the actual dissipation in most realistic systems. Nevertheless, we argue that the insights gained from studying the depolarizing noise channel are transferrable to those in generic noise channels.
To begin with, previous studies of the noisy quantum circuits have demonstrated that the exact form of the noise channel does not matter so long as the circuit itself is sufficiently random~\cite{schusterOperatorGrowthOpen2023a,angrisaniSimulatingQuantumCircuits2025}. The intuition is that sufficiently chaotic quantum channels would randomly rotate local spins on the Bloch sphere in an ergodic fashion, hence a noise channel would be equivalent to its rotation-average, which is the depolarization channel.

We expect a similar result to hold in our Lindbladian spectrum paradigm. In fact, as Eq.~\eqref{eq:lambda-as-trace} holds for generic noise channels, we have
\begin{equation}
\lambda = \sum_j \gamma_j \mathrm{tr}(V L_j^\dagger VL_j) -  \mathrm{tr}(V^2 L_j^\dagger L_j ) = -\frac{1}{2}\sum_j \gamma_j \mathrm{tr}\left([V,L_j]^\dagger [V,L_j]\right).
\end{equation}
As a crude argument, for an operator $V$ that is supported on $S$ sites, $[V,L_j]$ would only be non-zero for $S$ of the $j$'s, therefore $\lambda < O(\gamma S)$; on the other hand, a generic operator $V$ that acts on a site where $L_j$ lives would have an $O(1)$ commutator with $L_j$, leading to $\lambda \sim O(\gamma S)$. The case where $V$ is the sum of several small-sized operators follows similarly, since only operators with the same support have non-zero inner products, we can decompose the sum of $V$ and argue with each term separately.

As an example of non-depolarizing noise, consider the dephasing-decay noise, which includes both the dephasing channel with on-site jump operators $Z_j$, and the decay channel with on-site jump operators $S_j^-=\frac{1}{2}(X_j-iY_j)$. In Fig.~\ref{fig:S2}, we show that the first few low-lying eigenoperators with the dephasing-decay noise is largely the same as in the depolarizing noise for the TFIM and the MFIM, albeit to some degree with a less perfect correspondence compared to the depolarizing case. Importantly, for the depolarization channel, $\mathcal D$ is a Hermitian superoperator, leading to eigenoperators $V_i$ being orthogonal to each other up to first order in perturbation theory; this ceases to be true when decay noise is present. As can be seen in Fig.~\ref{fig:S2}(b), $V_1$ and $V_2$ are largely parallel to each other, so are $V_4$ and $V_5$. However, the fact that the first few eigenoperators are largely linear combinations of small-sized IOMs remain true.

However, exceptions exist where we cannot assume our Hamiltonian possesses the ability to average over local rotations. Roughly speaking, this happens when one operator is ``favored'' by both the noise channel and the Hamiltonian. For example, consider an XXX or XXZ Heisenberg spin chain subject to dephasing-decay noise. We observe that the projection operator $P_{\uparrow}$ to the totally polarized spin-up state $|\uparrow \dots \uparrow\rangle$ is an exact eigenoperator of such Lindbladians. In fact, due to spin conservation, $|\uparrow \dots \uparrow\rangle$ is an exact eigenstate of $H_\text{XXZ}$, hence $[H_\text{XXZ}, P_\uparrow]=0$. Furthermore, neither particle decay nor the phase-flip noise acts on the state $|\uparrow \dots \uparrow\rangle$: it is an eigenstate of $Z_j$, while $S_j^+$ annihilates the state (note: since we are considering $\mathcal L^\dagger$, ``acts on'' means the action of the dagger of the jump operator). Therefore, although $P_\uparrow$ is not a small-sized operator (its average size scales as $L$), it is a slow mode in the Lindbladian spectrum. Moreover, projections onto few-magnon states above this polarized state would also be slowly-decaying. Therefore, for Heisenberg spin chains with dephasing-decay noise, we will observe an additional set of (non-universal) slow modes as compared to in the depolarizing noise case.

\section{Other translational symmetry sectors}

\begin{figure}[!htbp]
    \centering
    \includegraphics[width=0.8\textwidth]{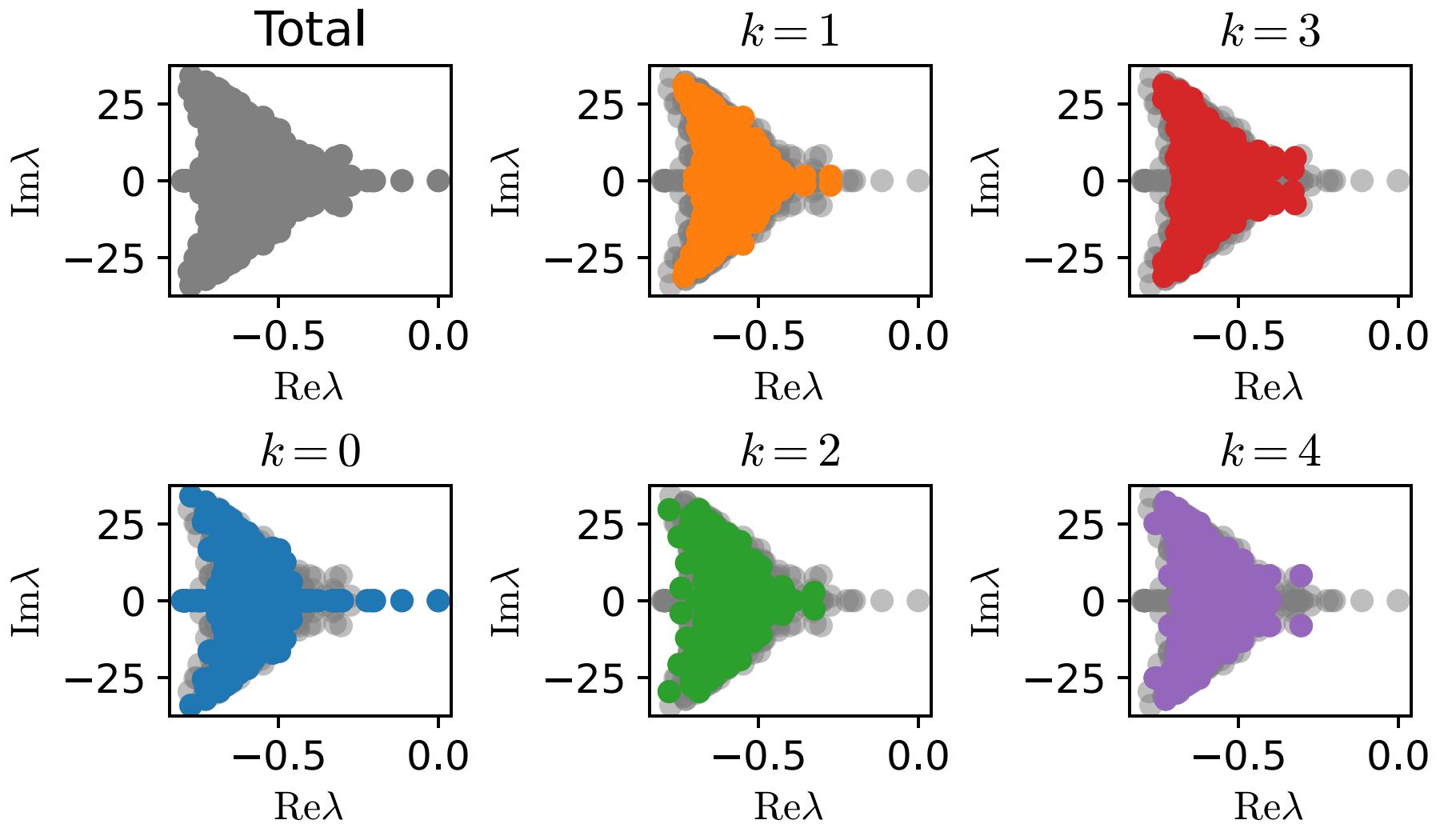}
    \caption{Lindbladian spectra for the TFIM with $J=1$ and $h=2$, noise channel being the depolarizing channel with $\gamma=0.1$, on a $L=8$ chain with PBC. The spectrum in all momentum sectors are shown. Here $k$ are presented in units of $\frac{2\pi}{L}$. We see that the spectra for different $k$'s roughly cover the same bulk region, while slow modes with small $|\re \lambda|$ exclusively lives in the $k=0$ sector.}
    \label{fig:S4}
\end{figure}

All the Lindbladian spectra presented in this work are for translationally invariant systems subject to periodic boundary conditions (PBCs) restricted to the zero-momentum sector. We exploit the translational invariance in order to push to larger system sizes. This shall be justified in most models, since we expect no slow modes to live in sectors with non-zero momenta. In fact, most known exact IOMs in PBC systems are translationally-invariant. The slowest modes in non-zero momentum sectors are expected to be the hydrodynamic modes $H_k=\sum e^{ikx} h_x$. As an approximate IOM, it has $\epsilon \sim k \sim \frac{2\pi}{L}$, which is pretty large for our system of $L\lesssim 10$. Therefore, we expect all the slow modes to lie in the zero-momentum sector. We exemplify this by presenting the Lindbladian spectrum of a TFIM in all momentum sectors in Fig.~\ref{fig:S4}

\section{Finite-size effects}

\begin{figure}[!htbp]
    \centering
    \includegraphics[width=0.8\textwidth]{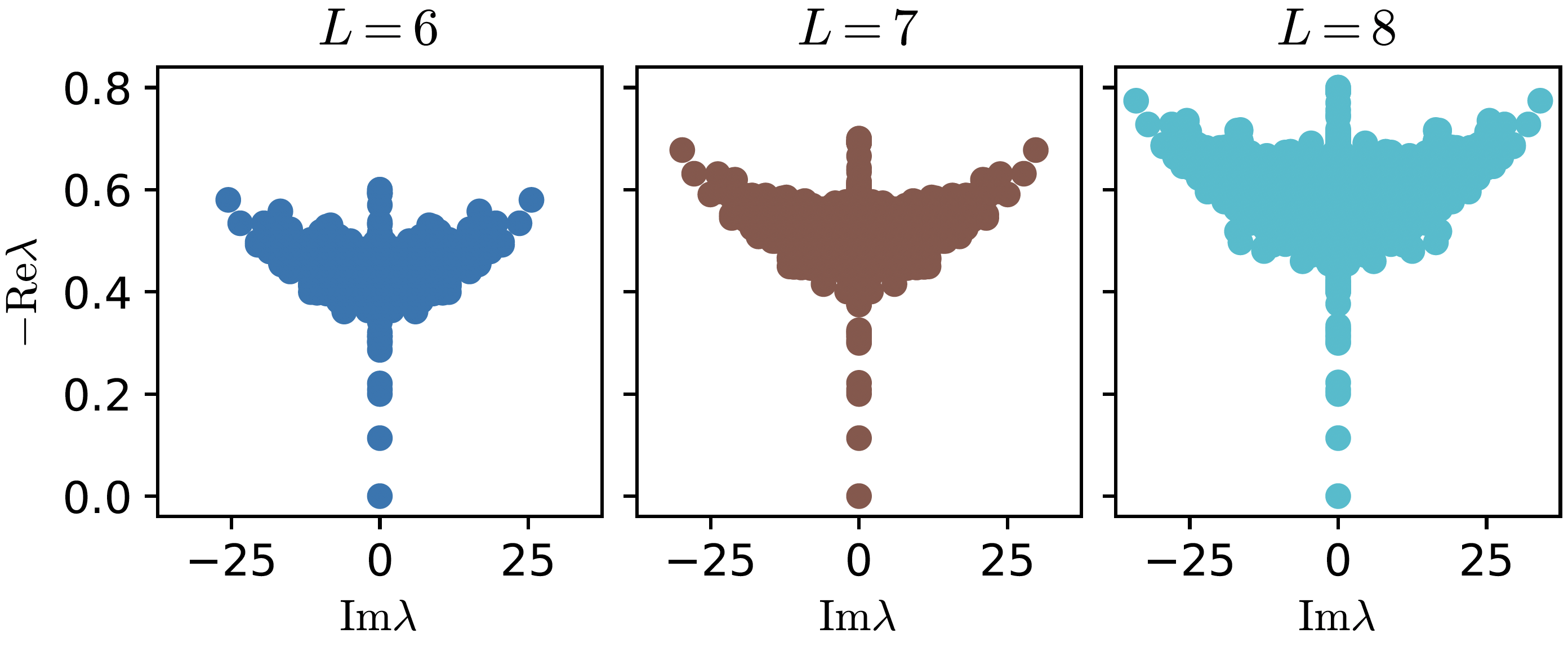}
    \caption{Lindbladian spectra for TFIMs with $J=1$ and $h=2$, noise channel being the depolarizing channel with $\gamma=0.1$, on chains with PBC in the zero-momentum sector. It can be observed that the bulk of the eigenvalues have $|\re\lambda|$ that becomes larger as $L$ is increased, while the slow modes' eigenvalues are largely insensitive to $L$.}
    \label{fig:SL}
\end{figure}

We briefly discuss how the fact that the system has finite size impacts the structure of the spectrum and our ability to identify slow modes. We have argued that IOMs appear as slow modes with a decay rate that is roughly equal to $\gamma$ times its size, and therefore does not scale with system size. By contrast, the generic eigenoperator of the Lindbladian is expected to spread out throughout the Hilbert space, and have an operator size that is proportional to $L$. This behavior is confirmed in Fig.~\ref{fig:SL}, where we observe that as we increase the system size, the ``bulk'' of the spectrum expands out to larger eigenvalues, while the slow-mode part doesn't undergo any noticeable change.

This structure reveals our ability to identify IOMs by diagonalizing Lindbladians on a finite-size system. Ideally, we know that the $|\re\lambda|$ of slow modes scale differently as that of fast modes; realistically, however, we can only know the exact value of $|\re\lambda|$ for a given $L$. Therefore, we can only observe a slow mode if it is not submerged by the bulk modes. For a given system size $L$, there is a threshold that is roughly proportional to $L$, such that IOMs with sizes larger than this threshold cannot be identified. For our numerics, this threshold ranges around 3 to 4.

\section{IOMs in the transverse-field Ising model}

The transverse-field Ising model is defined as
\begin{equation}
    H_\mathrm{TFIM} = \sum_i JZ_i Z_{i+1}+h X_i.
\end{equation}
It is known to be mappable to a free Majorana model with the Jordan-Wigner transformation. Let us define Majorana operators
\begin{align}
    \gamma_{i,A} &= \left[\prod_{j=1}^{i-1}X_j\right]Z_i, \\
    \gamma_{i,B} &= \left[\prod_{j=1}^{i-1}X_j\right]Y_i.
\end{align}
Notice that this definition on its own breaks the PBC, in the sense that $\gamma_{i+L,A(B)}\neq \gamma_{i,A(B)}$. However, the conserved quantities that we will later derive are valid in PBC systems as well. With the Majorana operators, we can write
\begin{equation}
    H_\mathrm{TFIM} = \sum_i i(J\gamma_{i,B}\gamma_{i+1,A}+h\gamma_{i,A}\gamma_{i,B}).
\end{equation}
This is a free (quadratic) model in terms of the Majorana operators. To find the IOMs, let us look at Majorana bilinear forms. Since $H$ is translationally invariant (for now we ignore the issue of periodicity raised by Majorana strings), we consider
\begin{equation}
Q = \sum_{x,y} (\gamma_{x,A},\gamma_{x,B}) q_{y-x} \begin{pmatrix}\gamma_{y,A} \\ \gamma_{y,B}\end{pmatrix}.
\end{equation}
Here, $q_{y-x}$ are $2\times 2$ matrices indexed by an integer. Using the commutation relation of Majorana operators, we obtain
\begin{equation}
\begin{cases}
    [H_\mathrm{TFIM},\gamma_{i,A}] = 2iJ\gamma_{i-1,B} - 2ih \gamma_{i,B}, \\
    [H_\mathrm{TFIM},\gamma_{i,B}] = -2iJ\gamma_{i+1,A} + 2ih \gamma_{i,A}.
    \end{cases}
\end{equation}
Further using $[A,BC]=[A,B]C+B[A,C]$, we have
\begin{multline}
[H_\mathrm{TFIM},Q] \\ = \sum_{x, y} \bigg[ (2iJ\gamma_{x-1,B} - 2ih \gamma_{x,B},-2iJ\gamma_{x+1,A} + 2ih \gamma_{x,A}) q_{y-x} \begin{pmatrix}\gamma_{y,A} \\ \gamma_{y,B}\end{pmatrix} + (\gamma_{x,A},\gamma_{x,B}) q_{y-x} \begin{pmatrix}2iJ\gamma_{y-1,B} - 2ih \gamma_{y,B} \\ -2iJ\gamma_{y+1,A} + 2ih \gamma_{y,A}\end{pmatrix} \bigg] \\
= \sum_{x,y} (\gamma_{x,A},\gamma_{x,B}) \bigg[
\begin{pmatrix}&2ih\\-2ih&\end{pmatrix}q_{y-x}
+ q_{y-x}\begin{pmatrix}&-2ih\\2ih&\end{pmatrix} 
+\begin{pmatrix}&0\\2iJ&\end{pmatrix}q_{y-x-1}
\\
+ \begin{pmatrix}&-2iJ\\0&\end{pmatrix}q_{y-x+1}
+q_{y-x-1}\begin{pmatrix}&0\\-2iJ&\end{pmatrix}
+q_{y-x+1}\begin{pmatrix}&2iJ\\0&\end{pmatrix}
\bigg]  \begin{pmatrix}\gamma_{y,A} \\ \gamma_{y,B}\end{pmatrix} .
\end{multline}
Notice that due to the anti-commutation relation between the Majorana operators, an operator $P=\sum_{x,y}(\gamma_{x,A},\gamma_{x,B}) p_{y-x} \begin{pmatrix}\gamma_{y,A} \\ \gamma_{y,B}\end{pmatrix}$ would equal to zero if the matrix $p_{y-x}$ satisfies $p_{y-x}=p_{x-y}^T$. Now let $R=[H_\mathrm{TFIM},Q]$, we would have
\begin{multline}
r_{y-x} = \begin{pmatrix}&2ih\\-2ih&\end{pmatrix}q_{y-x}
+ q_{y-x}\begin{pmatrix}&-2ih\\2ih&\end{pmatrix} 
+\begin{pmatrix}&0\\2iJ&\end{pmatrix}q_{y-x-1}
\\
+ \begin{pmatrix}&-2iJ\\0&\end{pmatrix}q_{y-x+1}
+q_{y-x-1}\begin{pmatrix}&0\\-2iJ&\end{pmatrix}
+q_{y-x+1}\begin{pmatrix}&2iJ\\0&\end{pmatrix}.
\end{multline}
Using Pauli matrices, this can be rewritten as
\begin{equation}
r_{y-x} = J[Y-iX,q_{y-x+1}] + J[Y+iX,q_{y-x-1}]-2h[Y,q_{y-x}].
\end{equation}
Without loss of generality, let $q_{y-x}=-q_{x-y}^T$. Then we see that
\begin{equation}
r_{x-y}^T = J[Y+iX,q_{x-y+1}^T] + J[Y-iX,q_{x-y-1}^T] - 2h[Y,q_{x-y}^T] = -r_{y-x}.
\end{equation}
Therefore, $r_{y-x}^T = r_{x-y}$ is equivalent to $r_{y-x}=0$. Let $q_{y-x} = q^I_{y-x} I + q^X_{y-x}X  + q^Y_{y-x}Y + q^Z_{y-x}Z$, we would get a set of equations,
\begin{align}
J(q^Z_{y-x+1} + q^Z_{y-x-1})-2hq^Z_{y-x} & = 0, \label{eq:q-eq-1}\\
q^{Z}_{y-x+1} - q^Z_{y-x-1} & = 0, \label{eq:q-eq-2} \\
J(q^X_{y-x+1} + q^X_{y-x-1}) + iJ(q^{Y}_{y-x+1} - q^Y_{y-x-1}) - 2h q^X_{y-x} & = 0. \label{eq:q-eq-3}
\end{align}
In the meantime, $q_{y-x}=-q_{x-y}^T$ requires
\begin{align}
q^I_{y-x} & = -q^I_{x-y}, \label{eq:q-sym-1}\\
q^Z_{y-x} & = -q^Z_{x-y},  \label{eq:q-sym-2}\\
q^X_{y-x} & = -q^X_{x-y}, \label{eq:q-sym-3} \\
q^Y_{y-x} & = q^Y_{x-y}.  \label{eq:q-sym-4}\\
\end{align}
We can observe that these equations can be decoupled into three sets: (i) those involving $q^I$, (ii) those involving $q^Z$, and (iii) those involving $q^X$ and $q^Y$. We will look for solutions in these three sectors separately.

\paragraph{$q^I$ Sector.} We notice that the only equation in this sector is Eq.~\eqref{eq:q-sym-1}. Therefore, any $q^I$ satisfying this symmetry constraint gives rise to an IOM. A linear basis of these IOMs is given by taking $A_n$ to be the operator with $q_n^I=1$. This gives
\begin{equation}
\sum_x \gamma_{x,A}\gamma_{x+n,A} + \gamma_{x,B}\gamma_{x+n,B} \propto  YX^{n-1}Z-ZX^{n-1}Y =: A_{n}.
\end{equation}

\paragraph{$q^Z$ Sector.} This sector has no IOMs, since Eq.~\eqref{eq:q-sym-2} gives $q^Z_0=q^Z_1+q^Z_{-1}=0$, and combining with Eq.~\eqref{eq:q-eq-2} gives $q^Z_{x-y}=0$ for all $x,y$.

\paragraph{$q^X$ and $q^Y$ Sector.} Define
\begin{align}
x_m & =h(q^X_m+iq_m^Y) - J(q_{m-1}^X-iq_{m-1}^Y), \\
y_m & = J(q^X_m+iq_m^Y) - h(q_{m-1}^X-iq_{m-1}^Y),
\end{align}
which turns Eq.~\eqref{eq:q-eq-3} to $x_m=y_{m+1}$. For $J\neq h$, $x_m$ and $y_m$ are independent variables, as is evident from the existence of the inversion relation,
\begin{align}
q_m^X & = \frac{J(y_m-x_{m+1})+h(y_{m+1}-x_m)}{2(J^2-h^2)}, \\
q_m^Y & = \frac{J(y_m+x_{m+1})-h(y_{m+1}+x_m)}{2i(J^2-h^2)}.
\end{align}
In this case, the space of IOMs are generated by a basis where each operator correspond to a single pair $(x_m,y_{m+1})$ being non-vanishing. With a bit of calculation, this leads to
\begin{multline}
\sum_x J\gamma_{x,B}\gamma_{x+m+1,A} + h\gamma_{x,A}\gamma_{x+m,B}-h\gamma_{x,B}\gamma_{x+m,A} - J \gamma_{x,A}\gamma_{x+m-1,B} \\
\propto
\begin{cases}
    JZX^{m}Z-hZX^{m-1}Z-hYX^{m-1}Y+JYX^{m-2}Y =: B_m ,& m\geq 2 ,\\
JZXZ - hZZ - hYY - JX := B_1, & m=1, \\
JZZ + hX:= B_0 = H, & m=0.
\end{cases}
\end{multline}
This gives the complete set of Majorana-bilinear IOMs of the TFIM. Note that as each of them is a translationally-invariant sum of local Pauli terms, they would not be subject to the sign problem of Majorana operators on the boundary.

\section{Approximate conserved quantity in the mixed-field Ising model}

In Fig.~{\figmfimaprx}(b) in the main text, we have mentioned that an approximate conserved operator exists for the mixed-field Ising model with $J=1$, $h_x=h_z=0.4$. This operator was constructed in Section V of Ref.~\cite{wurtzEmergentConservationLaws2020}. The idea is to take the domain-wall number operator,
\begin{equation}
N_0 = \sum_i Z_i Z_{i+1},
\end{equation}
which is also the $J$-part (hence dominant part) of the Hamiltonian, and perform a procedure akin to the Schrieffer-Wolff transformations to construct an operator that has a smaller commutator with the Hamiltonian. The explicit form of the resulting ``dressed particle number operator'' is
\begin{multline}
N = \sum_i 0.9530 Z_i Z_{i+1} + 0.2135 X_i + 0.1927 Z_i X_{i+1} Z_{i+2} + 0.0616 Z_i X_{i+1} X_{i+2} Z_{i+3} - 0.0398(Z_i X_{i+1} + Z_{i+1} X_i) \\
- 0.0243 Y_i Y_{i+1} + 0.0211 Z_i X_{i+1} X_{i+2} X_{i+3} Z_{i+4} - 0.0164(X_i X_{i+1} Z_{i+2} + Z_i X_{i+1} X_{i+2}) + 0.0098 Z_i + \ldots.
\end{multline}

\section{IOMs in the Heisenberg model}

The spin-1/2 Heisenberg spin chain is given by the Hamiltonian
\begin{equation}
H = \sum_i J_x X_i X_{i+1} + J_y Y_i Y_{i+1} + J_z Z_i Z_{i+1}.
\end{equation}
If $J_x=J_y$, it is called the XXZ model; if furthermore $J_z=J_x=J_y$, it is called the XXX model. For the XXZ model, we often define $\Delta=J_z/J_x$ as the anisotropy parameter.

Apart from the energy, we can obviously see that the total spin $S_z=\sum_i Z_i$ is conserved in the XXZ model, and $S_x=\sum_i X_i$ and $S_y=\sum_i Y_i$ are also conserved in the XXX model. Furthermore, the algebraic Bethe ansatz gives an infinite family of conserved quantities~\cite{grabowskiQUANTUMINTEGRALSMOTION1994}. Notably, the first non-trivial one is
\begin{equation}
K = \sum_i J^{-1}_x Z_i X_{i+1} Y_{i+2} + J^{-1}_y X_i Y_{i+1} Z_{i+2} + J^{-1}_z Y_i Z_{i+1} X_{i+2} - J^{-1}_x Y_i X_{i+1} Z_{i+2} - J^{-1}_y Z_i Y_{i+1} X_{i+2} - J^{-1}_z X_i Z_{i+1} Y_{i+2}.
\end{equation}
Furthermore, the model is known to have a family of quasi-local IOMs~\cite{ilievskiCompleteGeneralizedGibbs2015a,ilievskiQuasilocalChargesIntegrable2016a}. The detailed expression for the quasi-local IOMs are complicated, but they are known to reduce to the spin current in the $\Delta \to 0$ limit~\cite{pereiraExactlyConservedQuasilocal2014}.

\begin{figure}[!htbp]
    \centering
    \includegraphics[width=\textwidth]{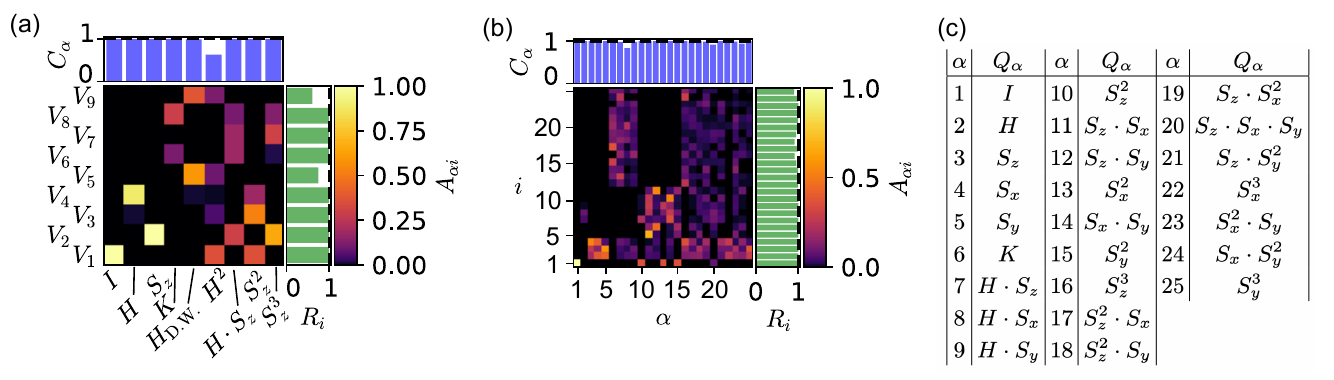}
    \caption{Additional results on Heisenberg spin chains. Numerics done on $L=9$ periodic boundary chains in the $k=0$ sector, with depolarizing noise of strength $\gamma=10^{-3}$. (a) The first few eigenoperators of the XXZ model with $\Delta=10$. (b, c) The first dozens of eigenoperators of the XXX model.}
    \label{fig:S1}
\end{figure}

In Fig.~\ref{fig:S1}, we present additional numerical results on the Heisenberg spin chains. In Fig.~\ref{fig:S1}(a), we present the spectrum of the XXZ model with $\Delta=10$. In addition to the usual IOMs, we found a slow mode roughly corresponding to the domain-wall swap operator, $H_\text{D.W.} = XX+YY+ZXXZ+ZYYZ$. As a physical interpretation, when $\Delta\gg 1$, the model is dominated by the domain-wall energy term $Z_iZ_{i+1}$. Meanwhile, we can rewrite the domain-wall swap operator as
\begin{equation}
H_\text{D.W.} \propto \sum_i  (S_{i+1}^+ S_{i+2}^- + S_{i+1}^- S_{i+2}^+) \delta_{Z_i,Z_{i+3}}.
\end{equation}
That is, it allows a quasiparticle to hop between the sites $i+1$ and $i+2$, given that the sites $i$ and $i+3$ are either both filled or both empty. We can observe that these are exactly the spin-exchange processes that happen within a sector with a fixed number of domain walls. This is similar to the emergent conserved quantities in prethermalizing systems~\cite{abaninRigorousTheoryManyBody2017a}.

In Fig.~\ref{fig:S1}(b,c), we present the eigenoperators of the XXX model. We find that up to 25 eigenoperators have perfect correspondence with the known IOMs. Notably, we are not able to observe the quasi-local IOM among them; this might be due to it having a size larger than 3, hence being overwhelmed by the finite-size effect.

\end{document}